\def\ref{\par\noindent\hang}

\def\spose#1{\hbox to 0pt{#1\hss}}
\def\approxlt{\mathrel{\spose{\lower 3pt\hbox{$\sim$}}
	\raise 2.0pt\hbox{$<$}}}
\def\approxgt{\mathrel{\spose{\lower 3pt\hbox{$\sim$}}
	\raise 2.0pt\hbox{$>$}}}

\def\multleft#1{\hbox to size{\vbox {\halign {\lft{##}\cr #1}}\hfill}\par}
\def\multright#1{\hbox to size{\vbox {\halign {\rt{##}\cr #1}}\hfill}\par}

\def\degmark{^\circ}
\def\today{\ifcase\month\or January\or February\or March\or April\or May\or
      June\or July\or August\or September\or October\or November\or December\fi
      \space\number\day, \number\year}
\def\$<${\thinspace}
\def\s{\hbox{\phantom{5}}}	

\def\boxit#1{\vbox{\hrule\hbox{\vrule\kern3pt\vbox{\kern3pt
          #1 \kern3pt}\kern3pt\vrule}\hrule}}

\def\cm{{\rm\thinspace cm}}

\def\erg{{\rm\thinspace erg}}

\def\keV{{\rm\thinspace keV}}

\def\Jy{{\rm\thinspace Jy}}

\def\Mpc{{\rm\thinspace Mpc}}
\def\Msun{\hbox{$\rm\thinspace M_{\odot}$}}

\def\s{{\rm\thinspace s}}

\def\ergpcmsqps{\hbox{$\erg\cm^{-2}\s^{-1}\,$}}

\def\ergps{\hbox{$\erg\s^{-1}\,$}}

\def\pcmsq{\hbox{$\cm^{-2}\,$}}

\def\ps{\hbox{$\s^{-1}\,$}}

\documentstyle[psfig]{mn}

\title{{\it ASCA} observations of the nearby galaxies Dwingeloo~1 and Maffei~1}

\author[C.~S.~Reynolds et al.]
{\parbox[]{6.5in}{C.~S.~Reynolds$^1$\thanks{Present address: JILA,
University of Colorado, Boulder, CO 80309-0440, USA.}, A.~J.~Loan$^1$, A.~C. Fabian$^1$, K.~Makishima$^2$, W.~N.~Brandt$^1$ and T.~Mizuno$^2$}\\
\\
$^1$Institute of Astronomy, Madingley Road, Cambridge CB3 OHA\\
$^2$Department of Physics, University of Tokyo, Hongo, Bunkyo-ku, Tokyo,
Japan}

\date{}

\begin{document}

\maketitle

\begin{abstract}
We present {\it ASCA} observations of the nearby galaxies Dwingeloo~1
(Dw1) and Maffei~1 (Mf1).  X-ray sources are clearly detected within 3
arcminutes of the optical nuclei of both galaxies.  Despite the low
Galactic latitude of these fields ($|b|<1\degmark$) we conclude, on
probability and spectral grounds, that the detected sources are
intrinsic to these galaxies rather than foreground or background
interlopers.  The Dw1 source, designated Dw1-X1, is interpreted as
being either a hyper-luminous X-ray binary (with a 0.5--10\,keV
luminosity of more than $10^{39}\ergps$) or an X-ray bright supernova.
The Mf1 emission is hard and extended, suggesting that it originates
from a population of X-ray binaries.  Prompted by the Dw1-X1 results,
we discuss the nature of hyper-luminous X-ray binary systems.  Such
sources are commonly seen in nearby galaxies with a frequency of
approximately one per galaxy.  We present a possible connection
between these luminous systems and Galactic superluminal sources.
\end{abstract}


\section{Introduction}

\begin{table}
\caption{Basic parameters for the {\it ASCA} observations of Dw1 and
Mf1.  The table shows a) the good SIS and GIS times (given the
selection criteria discussed in the text), b) total number of source
and background photons in the SIS and GIS source regions, c) the SIS
and GIS count rates of the detected sources (background subtracted),
d) observed (i.e. absorbed) 0.5--10\,keV and 2--10\,keV fluxes of the
sources and e) intrinsic (i.e. absorption corrected) 0.5--10\,keV and
2--10\,keV luminosities of the sources assuming them to lie at a
distance of 3\,Mpc (the assumed distance of Dw1 and Mf1).  The fluxes
and luminosities were calculated using the best fit power-law model with
cold absorption, although the values are fairly insensitive to
the exact model used.}
\begin{tabular}{lll}
& Dw1 & Mf1 \\\hline
Good SIS time (s) & 37462 & 35000 \\
Good GIS time (s) & 41508 & 39477 \\
SIS total counts & 472 & 1206 \\
GIS total counts & 1238 & 1957 \\
SIS count rate (cts\ps)& $8.7\times 10^{-3}$ & $2.4\times 10^{-2}$ \\
GIS count rate (cts\ps)& $1.1\times 10^{-2}$ & $2.4\times 10^{-2}$ \\
$F_{2-10\keV}(10^{-13}\ergpcmsqps)$ & 6.7 & 12\\
$F_{0.5-10\keV}(10^{-13}\ergpcmsqps)$ & 7.5 & 15\\ 
$L_{2-10\keV}(10^{38}\ergps)$ & 8.1 & 14 \\
$L_{0.5-10\keV}(10^{38}\ergps)$ & 14 & 25 \\\hline
\end{tabular}
\end{table}

It is now becoming apparent that `normal' galaxies host a variety of exotic
objects (Makishima 1994; Fabbiano 1995).  Although the Milky Way and the
Andromeda Galaxy (M31) are unusually quiescent in the X-ray band, recent
sensitive studies of other nearby spiral galaxies have revealed a number of
sources of high-energy radiation.  These are interpreted variously as
unusually powerful accreting stellar sources, supernova remnants,
superbubbles of hot gas associated with starbursts, or active galactic
nuclei (AGN) with unusually low X-ray luminosity.  Examples of
low-luminosity AGN (LLAGN) include M106 (Makishima et al. 1994), M81 (Petre
et al. 1993; Ishisaki et al. 1996), M51 (Ehle, Pietsch \& Beck 1995), and
IC342 (Fabbiano \& Trinchieri 1987; Bregman, Cox \& Tomisaka 1993;
Makishima 1994; Okada et al. 1994).  Many of these galaxies, as well as
NGC1313 (Petre et al. 1994; Colbert et al. 1995), M82 (Stocke, Wurtz \&
K\"{u}hr 1991) and others (see Table~9.2 of Fabbiano 1995), have variable
non-nuclear pointlike sources with 2--10\,keV isotropic luminosities up to
$\sim 10^{40}\ergps$.  If these sources are X-ray binaries (XRB), the
observed luminosity requires either the accretion to be significantly above
the Eddington limit or onto an implausibly massive black hole (with
$M\approxgt 100\Msun$: see discussion in Section 2.3.2), or for the
emission to be highly anisotropic.  In any case, this class of XRB
represents an unusual departure from most Galactic XRB.  With the
sensitivity of {\it ASCA}, a typical 40\,000\,s observation could detect
(at $5\sigma$ confidence) a $10^{40}\ergps$ point source at distances
$\approxlt 30\Mpc$.  Any detailed study would require a flux $\sim 2$
orders of magnitude larger than this detection threshold.  Thus, such
sources can only be studied in local galaxies (with distances $\approxlt
3\Mpc$).

In 1994 August the Dwingeloo Obscured Galaxy Survey detected the
21\,cm emission from a new nearby barred spiral galaxy -- now known as
Dwingeloo~1 (Dw1; Kraan-Korteweg et al. 1994).  Tully-Fisher distance
estimates place Dw1 at $\sim 3\Mpc$ (Loan et al. 1996).  Optical
photometry implies that Dw1 has an unusually blue core, suggesting
possible nuclear activity.  X-ray observations of Dw1 present us with
the ideal opportunity to study the high-energy emission from a
previously unknown, major, nearby spiral galaxy.

Maffei~1 (Mf1; Maffei 1968) is the nearest major elliptical galaxy, at
a distance of $\sim 3$\,Mpc (van den Bergh 1994).  An observation of
Mf1 with the imaging proportional counter (IPC) of the {\it Einstein}
observatory found a weak X-ray source with an observed 0.8--3.5\,keV
flux of $(3.3\pm 0.5)\times 10^{-13}\ergpcmsqps$, identifiable with
Mf1 (Markert \& Donahue 1985).  The alignment of several other sources
in the field with possible optical counterparts strongly suggested
that the astrometry was correct and that the identification with Mf1
was secure.  The observation hinted that this source was spatially
extended, although the possibility of foreground contamination and
blending of some nearby sources prevented a definitive statement from
being made.  Markert \& Donahue concluded that the X-ray emission from
Mf1 is from either a LLAGN, a centrally condensed population of XRB, or
hot interstellar medium (ISM).

Both Dw1 and Mf1 lie at low Galactic latitude ($b=-0.1\degmark$ and
$b=-0.6\degmark$ respectively).  {\sc H\,i} measurements and
extinction estimates yield neutral hydrogen column densities towards
Dw1 and Mf1 of $7\times 10^{21}\pcmsq$ (Burton et al. 1996) and
$8\times 10^{21}\pcmsq$ (Buta \& McCall 1983) respectively.  This high
level of absorption and extinction severely hampers studies of these
galaxies at optical wavelengths.  Infrared studies are difficult due
to confusion caused by foreground Galactic stars.  However,
observations in the hard X-ray band ($\approxgt 2\keV$) are little
affected by the low Galactic latitude.

Here we present {\it ASCA} observations of Dw1 and Mf1.  The range of
energies observable by the detectors on {\it ASCA}, 0.5--10\,keV, lies
mostly above the threshold at which Galactic absorption is important.
Coupled with the high sensitivity, superior spectral resolution and
imaging ability of {\it ASCA}, these observations constitute the most
detailed high-energy study to date of these two nearby galaxies.
Sections 2 and 3 present our analysis of the data for Dw1 and Mf1
respectively.  We clearly detect emission within a few arcmins of the
optical nucleus of each galaxy (1~arcmin corresponds to 1\,kpc at the
distance of Dw1 and Mf1).  We discuss physical interpretations for
these emissions.  The Dw1 source, designated Dw1-X1, is interpreted as
being either a hyper-luminous XRB or a X-ray bright
supernova.  The Mf1 emission is hard and extended, suggesting that it
originates from a population of X-ray binaries.  Prompted by our
Dw1-X1 results, we discuss the nature of hyper-luminous X-ray binary
systems.  Such sources are commonly seen in nearby galaxies with a
frequency of approximately one per galaxy (Fabbiano 1995).  In Section
4, we present a possible connection between these luminous systems and
Galactic superluminal sources.  Section 5 presents our conclusions.

\section{Dwingeloo~1}

\begin{figure}
\centerline{\psfig{figure=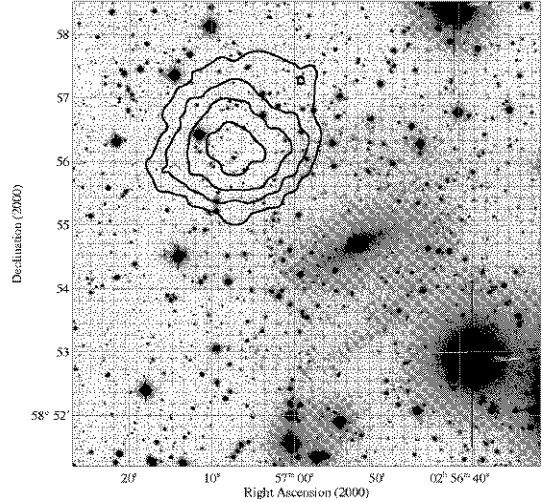,width=0.5\textwidth}}
\caption{$I$-band image of Dw1 taken with the INT.  The optical nucleus
of Dw1 lies slightly to the right of the frame centre.  Overlaid are
logarithmically spaced contours of the GIS2 full band image.  Dw1-X1
is clearly displaced $\sim 3$\,arcmin north-east of the optical nucleus of
Dw1.  Contours correspond to 0.89, 0.71, 0.56, 0.45 of the peak
X-ray surface brightness.   For comparison, the average background level
is $\sim 15$ per cent of the peak surface brightness with maximum
fluctuations up to $\sim 30$ per cent of the peak surface brightness.}
\end{figure}

On 1995 September 14/15, {\it ASCA} observed Dw1 for approximately one
day.  We used both Solid-state Imaging Spectrometers (SIS) in 1-CCD
mode (chip 1 for SIS0 and chip 3 for SIS1).  Further details of the
{\it ASCA} instrumentation can be found in Tanaka, Inoue \& Holt
(1994).  Data from the SIS were cleaned in order to remove the
effects of hot and flickering pixels and subjected to the following
data-selection criteria: i) the satellite should not be in the South
Atlantic Anomaly (SAA), ii) the object should be at least $5\degmark$
above the Earth's limb, iii) the object should be at least
$25\degmark$ above the day-time Earth limb and iv) the local
geomagnetic cut-off rigidity (COR) should be greater than 6\,GeV/$c$.
Data from the Gas Imaging Spectrometers (GIS) were cleaned to remove
the particle background and subjected to the following data-selection
criteria: i) the satellite should not be in the SAA, ii) the object
should be at least $7\degmark$ above the Earth's limb and iii) the COR
should be greater than 7\,GeV/$c$.  SIS and GIS data that satisfy
these criteria shall be referred to as `good' data.  Table~1 gives the
`good' integration times for the SIS and GIS.

\subsection{Image analysis}

\noindent Images were extracted from the good data for each of the
four instruments.  We clearly detected a bright point-like source
(i.e. with a surface brightness profile consistent with that of the
point spread function in both the SIS and GIS instruments).  Hereafter
this source will be referred to as Dw1-X1.  Figure~1 shows contours of
this X-ray image overlaid on an $I$-band image from the Isaac Newton
Telescope (INT) of this region (Loan et al. 1996).  Despite the
numerous Galactic stars, Dw1 can be clearly seen at the centre of
Fig.~1 displaying a strong bar and faint spiral arms (extending
outwards in a clockwise direction).  As reported in Table~1, the
2--10\,keV flux of this source is $6.7\times 10^{-13}\ergpcmsqps$
(assuming the best fit power-law spectrum of Section 2.2).  Dw1-X1 is
displaced $\sim 3$\,arcmins to the north-east of the optical nucleus of Dw1.
A recent study shows that the pointing accuracy of {\it ASCA} has a
random positional offset following a Poisson probability distribution
with mean of $\sim 0.4$\,arcmin (Gotthelf 1996).  Thus Dw1-X1 is
extremely unlikely to be associated with the optical nucleus of Dw1
(with a probability of $\sim 5\times 10^{-4}$).  We note that the {\sc
H\,i} disk of Dw1 has a diameter $\sim 12\arcmin$ and so easily
encompasses the position of Dw1-X1 (Burton et al. 1996).  No nuclear
emission is seen from Dw1: the corresponding limit on the 2--10\,keV
flux of the nucleus is $F_{2-10}\approxlt 1\times
10^{-13}\ergpcmsqps$.

\subsection{Spectral and temporal analysis of Dw1-X1}

\begin{figure}
\centerline{\psfig{figure=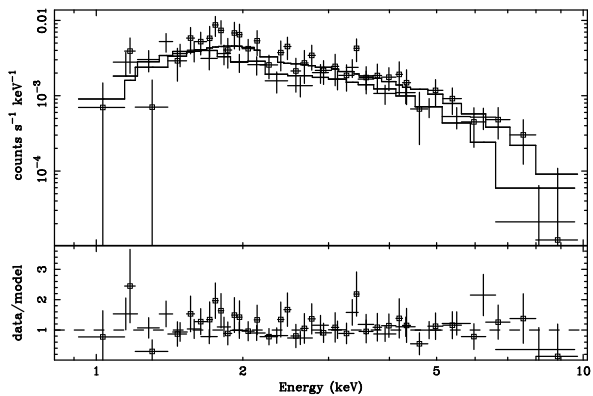,width=0.5\textwidth,angle=270}}
\caption{SIS0 (plain crosses) and GIS2 (crosses with open circles)
spectra of Dw1-X1.  Although all four instruments were used for
spectral fitting, only these two are displayed in order to maintain
clarity.  The top panel shows the data fitted with the best fit
absorbed power-law model of Section 2.2 (shown by solid lines).  The
bottom panel illustrates the ratio of data to this best fit model.}
\end{figure}

Light curves and spectra were extracted for the counts contained
within a circular region centred on Dw1-X1 with a radius of 4\,arcmin.
Background spectra were also extracted from source free regions of the
same field of view for each instrument.  For the purposes of
spectroscopy, only 0.6--10\,keV SIS data and 0.9--10\,keV GIS data
were used.  The spectra were binned so as to have at least 20 photons
per energy bin: this  requirement ensures that $\chi^2$ statistics
apply.  Table~1 shows the count rates, 2--10\,keV flux and the
2--10\,keV luminosity if we choose to place it at the distance of Dw1.
The light curves of all instruments are consistent with the source
being constant ($\chi^2/$dof=21/33 for the SIS0 light curve) although
variability would have to be dramatic in order to be detectable in a
source with such a low count rate.  We estimate that a variation by a factor
of two sustained over $\approxgt 2000$\,s (corresponding to the good
data period of one or more {\it ASCA} orbits) would be marginally
detectable at the 3-$\sigma$ level.

\begin{table*}
\begin{center}
\caption{Spectral fitting results for Dw1-X1.  Abbreviations are as follows:
bb=black-body; brems=thermal bremsstrahlung; rs=Raymond-Smith thermal
plasma code (Raymond \& Smith 1977); pl=power-law; NH=cold absorbing
column. All errors are quoted at the 90 per cent confidence level for one
interesting parameter, $\Delta\chi^2=2.7$. }
\begin{tabular}{llcc}
Model 		& parameter 		& value 	& $\chi^2$/dof \\\hline
bb+NH		& temperature		& $kT=1.03^{+0.09}_{-0.08}\keV$ & 137/138 \\
		& cold absorbing column & $N_{\rm H}<3\times 10^{21}\pcmsq$ & \\\hline
bb+bb+NH	& temperatures		& $kT_1=0.49^{+0.18}_{-0.15}\keV$ & 117/136 \\
		&			& $kT_2=1.5^{+0.5}_{-0.3}\keV$ & \\
		& rel. norms at 1\,keV  & $A_1/A_2=0.64$ & \\
		& cold absorbing column & $N_{\rm H}=(1.0\pm 0.6)\times 10^{22}\pcmsq$ & \\\hline
brems+NH 	& temperature		& $kT=6.6^{+4.0}_{-2.1}\keV$ & 120/138 \\
		& cold absorbing column	& $N_{\rm H}=(1.1\pm 0.3)\times 10^{22}\pcmsq$ & \\\hline
rs+NH		& temperature		& $kT=5.9^{+4.3}_{-1.8}\keV$ & 120/137 \\
		& abundance		& $Z<0.54$ & \\
		& cold absorbing column	& $N_{\rm H}=(1.1\pm 0.3)\times 10^{22}\pcmsq$ & \\\hline
pl+NH		& photon index		& $\Gamma=2.0\pm 0.3$ & 119/138 \\
		& cold absorbing column	& $N_{\rm H}=1.4^{+0.4}_{-0.3}\times 10^{22}\pcmsq$ & \\\hline
\end{tabular}
\end{center}
\end{table*}

Spectral fitting was performed using background subtracted data from all
four {\it ASCA} instruments simultaneously.  We initially considered four
spectral models: black-body emission, thermal bremsstrahlung, Raymond-Smith
thermal plasma emission (including line emission) and a power-law.  The
effects of cold absorption were included, leaving the column density as a
free parameter.  There are known small discrepancies (at the 10--20 per
cent level) between the normalisation of each of the four instruments.
Thus, the best fit spectrum was allowed to have independent normalisations
in each instrument.  The results of these fits are shown in Table~2.  

The bremsstrahlung, Raymond-Smith and power-law models all describe the
data equally well.  The black-body model is a poorer description of the
data and results in an unmodelled hard tail to the spectrum.  Adding a
second black-body component leads to a significant improvement in the
goodness of fit ($\Delta\chi^2=20$ for extra 2 parameters), as can be seen
from Table~2.  Note that adding a power-law, bremsstrahlung or
Raymond-Smith component to the single black-body model in an attempt to
model the hard tail leads to that component completely dominating the
spectrum (i.e. the models essentially become one of those listed in
Table~2).  Figure~2 shows the best-fitting (folded) power-law model
compared to the data together with a ratio plot of the data divided by the
model.

A wide variety of astrophysical X-ray sources display iron K$\alpha$
line emission (with energies ranging from 6.4--6.9\,keV depending on
the dominant ionization state of iron).  Thus, although the simple
models presented above are completely adequate descriptions
of the {\it ASCA} spectrum of Dw1-X1, it is instructive to find an upper
limit to the equivalent width (EW) of any iron line emission.  A narrow
line at 6.4\,keV (representing cold iron) was added to both the
thermal and power-law models above.  The addition of the iron line produced
no improvement in the goodness of fit:  the best fit EW is zero.  For
all of the above models, an upper limit can be set of EW$<$600\,eV (at the
90 per cent confidence level for one interesting parameter,
$\Delta\chi^2=2.7$).

\subsection{The physical nature of Dw1-X1}

\subsubsection{Locating the source along our line of sight}

Any discussion of the nature of Dw1-X1 must begin by considering its
location along our line of sight.  To demonstrate that this source is
intrinsic to Dw1, we must exclude the possibilities of it
being a foreground Galactic source (e.g. a flare star or XRB) or a
background AGN.

The spectral results of Section 2.2 clearly show the effects of soft X-ray
absorption by cold material.  For the power-law, bremsstrahlung and
Raymond-Smith models presented above, the column density inferred from
X-ray absorption exceeds the Galactic column inferred from {\sc H\,i} and
extinction measurements ($N_{\rm H}\approx 7\times 10^{21}\pcmsq$).  The
double black-body model admits a rather lower cold column due to the
intrinsic low-energy turnover of the model.  However, even within this
model, the 90 per cent lower limit to the cold column is $N_{\rm H}>4\times
10^{21}\pcmsq$ and the preferred column is $N_{\rm H}=1\times
10^{22}\pcmsq$.  

The fact that the spectrum of Dw1-X1 seems to suffer absorption by at least
a significant fraction of the Galactic suggests that Dw1-X1 lies no closer
than the far side of our Galactic disk and is probably extragalactic.  If
the power-law or optically-thin thermal models are a correct description of
the spectrum, there is also evidence for intrinsic absorption in the source
(above that expected from the Galaxy).  However, self-absorption of the
21-cm line can lead to {\sc H\,i} measurements underestimating the true
{\sc H\,i} column density when it is large.  This may allow the Galactic
{\sc H\,i} measurements to be consistent with the {\it ASCA} measurements.
We note that the above argument is based on our one-component spectral
fits.  

We can also argue against Dw1-X1 being of Galactic origin on
probabilistic grounds.  The question is precisely phrased as follows:
what is the probability of finding a Galactic X-ray source with at
least the flux of the {\it ASCA} Dw1 source within 3\,arcmin of the
optical nucleus of Dw1?  The {\it Einstein} Galactic Plane Survey
(Hertz \& Grindlay 1984) find a total of $\sim 0.4$ sources per square
degree with a 0.1--4.5\,keV flux above that of Dw1-X1 ($4\times
10^{-13}\ergpcmsqps$).  Most of these are Galactic sources.
Inspection of the {\it ROSAT} All Sky Survey (Voges et al. 1996) shows Dw1 to
lie in a representative region of the Galactic plane.  Given this the
probability of Dw1-X1 being a foreground source is $\sim 3\times
10^{-3}$.  This probability is an overestimate since it does not take
into account the spectral information discussed in the previous
paragraph (which implies it must lie at least on the far side of the
Galactic disk).  Thus, we discount this possibility.

Similarly, we use probabilistic arguments to address whether this
source can be a background AGN.  Again, the question is precisely
phrased as follows: what is the probability of finding an AGN with at
least the flux of the Dw1 source within 3\,arcmin of the optical
nucleus of Dw1?  The Deep {\it ROSAT\/} Survey of Boyle et al. (1994)
provides the most pertinent data with which to answer this question.
Using the best fit power-law model of Section 2.2, the unabsorbed
0.5--2\,keV flux of the Dw1 source is $5.4\times 10^{-13}\ergpcmsqps$.
The Deep {\it ROSAT\/} Survey suggests $\sim 0.5$ AGN per square
degree of this flux level or greater.  Assuming this to be
representative of the AGN background in the Dw1 region implies that
the probability of the Dw1 {\it ASCA} source being a background AGN is
only $\sim 4\times 10^{-3}$.  Thus, we also reject this possibility on
probabilistic grounds.

We conclude that the positional coincidence of Dw1-X1 with Dw1,
coupled with an X-ray spectrum that has most likely suffered Galactic
absorption, presents a strong argument for the source to be intrinsic
to Dw1.  Table~1 gives the unabsorbed luminosities of the source in
the 0.5--10\,keV and 2--10\,keV bands if it is placed at the distance
of Dw1 ($\sim 3$\,Mpc).  The 3\,arcmin offset from the optical nucleus
implies Dw1-X1 lies a distance of 2.6\,kpc from the centre of Dw1.
Although Dw1-X1 appears to lie beyond the optical extent of Dw1 (see
Fig.~1), this is a consequence of the high optical extinction along
this line of sight: the observed offset is completely consistent with
Dw1-X1 lying within Dw1.  Note for comparison that the {\sc H\,i}
extent of Dw1 is $\sim 12$\,arcmins corresponding to a diameter of
10\,kpc.
 
\subsubsection{The nature of the source}

The location of Dw1-X1 within the host galaxy suggests that it is a
stellar phenomenon.  Emission from a population of ordinary stars
would produce a much fainter and more diffuse emission than is
observed.  Thus, we are led to consider stellar exotica such as
superbubble regions (associated with starbursts), supernova remnants
and XRBs.

Large regions of hot gas associated with starburst activity, so-called
superbubbles, have probably been detected in several nearby galaxies (see
NGC~5408: Fabian \& Ward 1993; NGC~1672: Brandt, Halpern \& Iwasawa 1996).
The known examples tend to have soft optically-thin thermal X-ray spectra
(with $kT\sim 0.5\keV$).  Such emission is much softer than that of Dw1-X1.
Thus we confidently reject the hypothesis that Dw1-X1 is a superbubble.

A young supernova remnant that is expanding in a dense circumstellar medium
is a possibility for the observed emission (see Schlegel 1996 and detection
of SN1988Z in X-rays by Fabian \& Terlevich 1996).  Indeed, the spectrum
and luminosity of Dw1-X1 are similar to that of the X-ray bright supernova
SN1978K seen in NGC1313 (Ryder et al. 1993; Petre et al. 1994; Colbert et
al. 1995).  Since Dw1 has only been recently discovered, little direct
information exists on the supernova history of this galaxy.  However, we
can argue against the supernova hypothesis for Dw1-X1 on the basis of radio
observations.  X-ray bright supernovae are known to be strong sources of
continuum radio emission, e.g. SN1986J (Weiler, Panagia \& Sramek 1990) and
SN1988Z (Van Dyk et al. 1993).  Although very dependent on supernova age
and environment, we would expect any such supernova in Dw1 to have a radio
flux of $\sim 1\Jy$ at 1.4\,GHz (corresponding to $\nu L_\nu \approx
10^{37}\ergps$ at 1.4\,GHz, assuming a distance of $3\Mpc$).  Published
observations provide upper limits on the radio continuum luminosity of
Dw1-X1 of 50\,mJy at 408\,MHz (Green 1989) and 0.4\,mJy at 1.420\,GHz
(Burton et al. 1996).

Given the ambiguities mentioned above, the present observations cannot
rule out the possibility that Dw1-X1 is a young supernova remnant.
However, such X-ray bright supernovae are rare phenomena.  Moreover,
many other well studied galaxies display luminous non-nuclear sources
which do not correspond with the position of any known supernovae.
Several of these luminous non-nuclear sources are also known to be
variable: the luminous off-nuclear source associated with IC342 shows
clear short term variability (Makishima 1994; Okada et al. 1994).  For
these reasons, such sources are identified as accreting compact
objects.

The absorption corrected 0.5--$10\keV$ luminosity of Dw1-X1 (assuming
a power-law for for the spectrum) is $1.4\times 10^{39}\ergps$.
Depending on the details of the source, there is likely to be
significant emission outside of the {\it ASCA} band.  Including
emission from all other wavebands, the bolometric luminosity of this
source is probably as high as $\sim 3\times 10^{39}\ergps$.  This
luminosity exceeds any persistent Galactic XRB by a factor of a few.
Indeed, it corresponds to the Eddington limit for a $20\Msun$ compact
object.  This is in excess of the mass limit for a neutron star ($\sim
3\Msun$) or a black hole formed by the collapse of a normal massive
star (thought to be 10--$15\Msun$: e.g. see Timmes, Woosley \& Weaver
1996).

There are several ways to explain the luminosity of sources such as
Dw1-X1 in terms of accreting compact objects.  First, accretion onto a
stellar black hole at a rate in excess of the Eddington limit would
explain these sources.  Strong magnetic fields may channel material
onto the accreting object thereby allowing the Eddington limit to be
breached by a factor few (e.g. A0538$-$66; Charles et al. 1983 and
references therein).  If Dw1-X1 does indeed contain a massive stellar
black hole (with $M\sim 10-20\Msun$), only moderate super-Eddington
accretion is required in order to explain the measured luminosity.
However, there are many examples of more powerful ($L\approxgt
10^{40}\ergps$), but otherwise similar, sources in other nearby
galaxies (Fabbiano 1995).  To explain these as super-Eddington
accretion onto such black holes would require the Eddington limit to
be exceeded by a factor of 5--10.  It is difficult to understand how
such extreme super-Eddington accretion could occur in a stable way.

\begin{figure}
\centerline{\psfig{figure=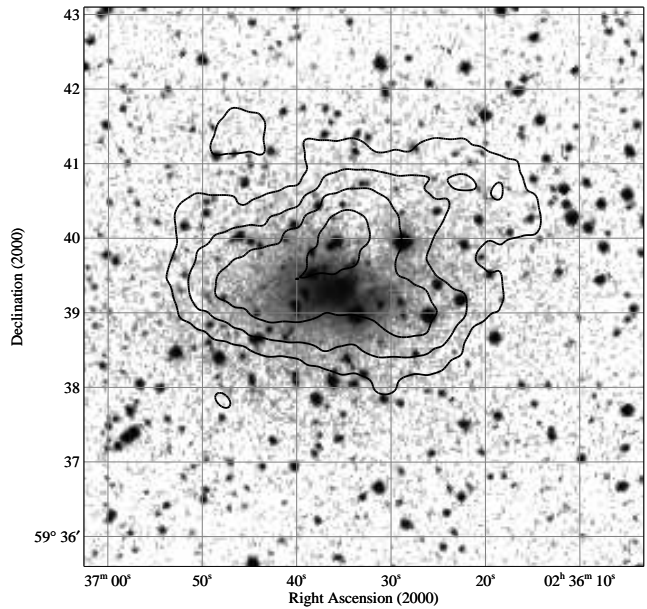,width=0.5\textwidth}}
\caption{Red ($E$-band) POSS plate image of Mf1 overlaid with contours of the
full-band GIS2 image.  Contours correspond to 0.93, 0.74, 0.59 and 0.47 of
the peak surface brightness.  For comparison, the average background level
is $\sim 10 per cent$ of the peak surface brightness with maximum
fluctuations up to $\sim 25$ per cent of the peak surface brightness.  The
small apparent offset between the X-ray source and optical galaxy is
interpreted as being due to positional uncertainty in the {\it ASCA}
pointing (see discussion in the text).  The source is significantly
extended and appears inhomogeneous.}
\end{figure}

Secondly, these luminous sources may represent Eddington or
sub-Eddington accretion onto a massive black hole (with $M\sim
100\Msun$ required in order to explain the most luminous examples).
However, it is difficult to explain the formation of a
$100\Msun$ non-nuclear black hole through the known processes of
stellar evolution\footnote{A black hole accreting at the Eddington
luminosity with an efficiency of $\eta\sim 0.1$ will have a growth
timescale of $\sim 10^8$\,yrs.  A binary system containing two massive
stars will evolve (and presumably lose much of its mass) on a much
more rapid timescale ($\sim 10^6$\,yrs).  Thus, it is difficult to
understand how a stellar black hole (with an initial mass of $\sim
10\Msun$) accreting at near the Eddington rate could be supplied with
fuel for a sufficient length of time in order to grow to $\sim
100\Msun$, unless the radiative efficiency is very low.}.  Of course, AGN
are thought to possess much more massive black holes: however, the
formation of such supermassive black holes is likely to be intimately
linked to their special location in the galaxy (e.g. see Rees 1984), and so
probably will not provide a role model for the formation of non-nuclear
massive black holes.

Finally, Dw1-X1 and similar luminous sources may be accreting stellar
objects (either neutron stars or black holes) from which the emission is
highly beamed.  This possibility is discussed in more detail in Section 4.

\section{Maffei~1}

{\it ASCA} observed Mf1 on 1995 September 11/12 for approximately one
day. Both SIS were used in 2-CCD mode (with the source positioned in
chip 1 in SIS0 and chip 3 in SIS1).  The data were cleaned and
filtered as described in Section 2.  The resulting good integration
times are given in Table~1.

\subsection{Image analysis}

\noindent Images were extracted from the good data for each of the four
instruments.  Figure~3 shows contours of the GIS2 image overlaid on
the POSS plate image of the Mf1 region.  An X-ray source is clearly
detected with an apparent offset of $\sim 30$\,arcsec from the optical
nucleus of Mf1.  Given the positional accuracy of {\it ASCA} (Gotthelf
1995; also see Section 2.1), this is consistent with the source being
centred on the optical galaxy (there is a 30 per cent probability of
obtaining a 0.5\,arcmin offset or more).  Comparing Fig.~3 with the
point source of Fig.~1 suggests that the Mf1 source has significant
extent and substructure.  In fact, both the hard-band image
($>2.5$\,keV) and soft-band image ($<2.5$\,keV) are {\it inconsistent}
with the point spread function.

\subsection{Spectral and temporal analysis of Maffei~1}

\noindent Light curves and spectra were 
extracted for the counts contained within a circular region centred on
the Mf1 source of radius 4\,arcmins.  Background spectra were
extracted from source free regions of the same field of view for each
detector.  The count rates, 2--10\,keV flux and 2--10\,keV luminosity
are reported in Table~1.  There is marginal evidence for variability
from the SIS light curves: the SIS light curves are inconsistent with
a constant model at the 90 per cent level ($\chi^2$/dof=41/31 and
$\chi^2$/dof=46/31 for constant model fits to SIS0 and SIS1
respectively).  However, no clear variations are correlated between
the two SIS instruments and no variability is evident in either GIS.
Thus, we cannot confidently claim to have detected real variability.

\begin{figure}
\centerline{\psfig{figure=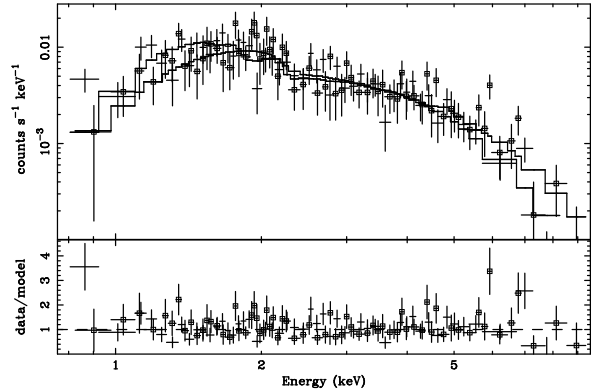,width=0.5\textwidth,angle=270}}
\caption{SIS0 (plain crosses) and GIS2 (crosses with open circles)
spectra of the Maffei~1 emission.  Although all four instruments were
used for spectral fitting, only these two are displayed in order to
maintain clarity.  The top panel shows the data fitted with the best
fit absorbed power-law model of Section 3.2 (shown by solid lines).
The bottom panel illustrates the ratio of data to this best fit model.}
\end{figure}

\begin{table*}
\begin{center}
\caption{Spectral fitting results for the Mf1 source.  Abbreviations are as follows:
brems=thermal bremsstrahlung; rs=Raymond-Smith thermal plasma code
(Raymond \& Smith 1977); pl=power-law; NH=cold absorbing column. All
errors are quoted at the 90 per cent confidence level for one
interesting parameter, $\Delta\chi^2=2.7$. }
\begin{tabular}{llcc}
Model 		& parameter 		& value 	& $\chi^2$/dof \\\hline
bb+NH		& temperature		& $kT=1.06^{+0.04}_{-0.05}\keV$ & 264/256 \\
		& cold absorbing column & $N_{\rm H}<8.5\times 10^{21}\pcmsq$ & \\\hline
brems+NH 	& temperature		& $kT=8.2^{+3.3}_{-2.0}\keV$ & 256/256 \\
		& cold absorbing column	& $N_{\rm H}=8.5^{+1.6}_{-1.4}\times 10^{21}\pcmsq$ & \\\hline
rs+NH		& temperature		& $kT=8.5^{+3.3}_{-2.0}\keV$ & 256/255 \\
		& abundance		& $Z<0.22$ & \\
		& cold absorbing column	& $N_{\rm H}=8.5^{+1.6}_{-1.4}\times 10^{21}\pcmsq$ & \\\hline
pl+NH		& photon index		& $\Gamma=1.80^{+0.17}_{-0.15}$ & 262/256 \\
		& cold absorbing column	& $N_{\rm H}=(1.0\pm 0.2)\times 10^{22}\pcmsq$ & \\\hline
\end{tabular}
\end{center}
\end{table*}

Spectral fitting was performed as in Section 2.2, using background
subtracted data from all four instruments simultaneously.  The results
of these fits are shown in Table~3.  All models
provide a satisfactory description of the data.  Figure 4 shows the
SIS0 and GIS2 spectra for Mf1 compared with the best fitting power-law
spectrum.

\subsection{The physical nature of the Maffei~1 emission}

Probabilistic arguments identical to those presented in Section 2.3.1 lead
us to conclude that the emission seen in this observation is intrinsic to
Mf1.  In fact, the probabilistic arguments are stronger in the case of Mf1
because the X-ray emission has a smaller angular separation from the
optical nucleus.  Given this conclusion, and the astrometric results of
Markert \& Donahue (1985), it is likely that the 0.5\,arcmin offset of the
centroid of the X-ray emission from the optical nucleus of the galaxy is
purely due to astrometric error within the {\it ASCA} observation.
Depending on the spectral model used, there is also spectral evidence for
the Galactic absorption expected if this source were extragalactic
(although spectral fitting with the black-body model can admit a rather
lower cold column due to the intrinsic low-energy turnover of the
black-body curve.)

To summarize the observed properties of the emission (given its
identification with Mf1), it has a 0.5--10\,keV luminosity of $2.5\times
10^{39}\ergps$ (assuming a power-law model) and a hard spectrum (well
characterised by either a power-law with $\Gamma=1.8^{+0.17}_{-0.15}$ or an
optically-thin thermal model with $kT=8.2^{+3.3}_{-2.0}\keV$) It is also
spatially extended in both the hard and soft bands (with some suggestion of
substructure within the image).  [We do not consider the black-body model
any further since it has no physical basis within the context of extended
emission from elliptical galaxies.]

Given the high Galactic column density to this source, we must address the
possibility that the observed extended structure is due to a Galactic dust
scattering halo around an otherwise point source of emission (e.g. see
Martin 1978).  The process of X-ray scattering by dust is highly energy
dependent (with scattering cross section varying roughly as $E^{-2}$, where
$E$ is the energy of the scattered photon).  Thus the fact that both hard
and soft band images are extended tends to rule out there being a
significant dust halo.  Additionally, Dw1-X1 has a similar line of sight
Galactic column yet displays a pointlike source of emission.  We conclude
that the effects of Galactic X-ray dust scattering on the observed extent
of the Mf1 emission is negligible.

This combination of properties places severe constraints on possible
explanations of this emission.  The imaging data alone might be
interpreted as evidence for emission from an extended hot ISM.
However, our spectral data discounts this possibility in the following
manner.  A hot ISM would produce thermal emission at the virial
temperature of the galaxy, $kT\approx 1\keV$.  The observed spectrum
is too hard for it to be dominated by such a component.  In fact,
adding such a component to either the power-law or thermal model
continuum models allows an upper limit to be placed on the ISM
emission (assuming it to be at 1\,keV and not to suffer intrinsic
absorption): the resulting upper limit on the 0.5--10\,keV luminosity
of the ISM (at the 90 per cent confidence level for one interesting
parameter, $\Delta\chi^2=2.7$) is $1.4\times 10^{38}\ergps$.  Thus,
the spectral constraints require such ISM emission to be less than 5
per cent of the total {\it ASCA} band flux.  This upper limit on the
ISM emission is low for a typical elliptical galaxy.  A contributing
factor may be the restricted region (corresponding to a radius of
3\,kpc at the distance of Mf1) that we choose for our analysis: there
could be a more spatially extended soft component which is
undetectable above the background due to Galactic absorption.  Also,
the upper limit is much less restrictive if a cooler ISM is considered
since most of the thermal ISM flux is then at energies which are
almost completely absorbed by the line of sight cold material.
Assuming a temperature of $kT=0.5\keV$, the upper limit on the
0.5--10\,keV luminosity becomes $2.2\times 10^{39}\ergps$.  Note that
if such a soft component was indeed present, a significantly larger
absorbing column than that reported in Table~3 would be necessary.

The extended and hard nature of the source suggests that the emission
is from a number of accreting compact objects.  Similar hard emission
has been found in other elliptical galaxies (Canizares, Fabbiano \&
Trinchieri 1987; Matsushita et al. 1994).  There are two obvious
possibilities to explain this emission.  First, the emission may be
due to a large number of low-mass X-ray binaries (LMXB).  Assuming a
typical LMXB 2--10\,keV luminosity of $\sim 10^{37}\ergps$,
approximately 100 such systems are required to provide all of the
observed emission.  Secondly, a large portion of the emission may be
due to a small number of luminous systems, similar to Dw1-X1.  There
may also be a contribution from a LLAGN.  Note that the spectra of
LLAGN, a population of LMXB and luminous compact objects are similar
(Makishima et al. 1989).  Thus, it is difficult to distinguish between
these various explanations on spectral grounds alone.

In order to make further progress with the study of the Mf1 emission,
high resolution X-ray imaging data of Mf1 is required in order to
investigate any substructure within the emission and, if possible,
resolve individual luminous sources.  Variability, or lack thereof, is
also an important diagnostic that should be addressed in future
observations.  We are engaged in a programme to make such observations.

\section{Beamed sources and Galactic superluminals}

X-ray sources producing $\sim 10^{40}\ergps$ appear to be relatively
common in nearby galaxies (Fabbiano 1995; Marston et al. 1995).  The
available numbers indicate a mean of approximately one such source per
galaxy.  One candidate class of objects that we propose here
encompasses the Galactic superluminals such as GRS~1915+105 and
GRO~J1655$-$40 (Mirabel \& Rodriguez 1995; Mirabel \& Rodriguez 1994;
Harmon et al. 1995; Hjellming \& Rupen 1995).  These objects are
strong, sometimes transient, X-ray sources which have radio jets
showing superluminal motion.  There is also a class of non-transient
Galactic sources (e.g. 1E~1740.7-2942; Mirabel et al. 1992) with jets
which have yet to show superluminal motion.  The velocity of the jets
in the superluminal objects is close to 0.92$c$.  Neither of these
objects has an X-ray luminosity much exceeding $10^{38}\ergps$, {\it
but this could be exceeded if we were to view the object down the
jet}.

Given a jet velocity of 0.92$c$ the corresponding Lorentz factor is
$\gamma\sim 3$, so the jet emission is beamed into a solid angle of
$\gamma^{-2}\sim 0.1$\,sr.  Thus the apparent luminosity could be $2\pi
\gamma^2\sim 60$ times higher along that direction than the actual power of
the jet (allowing for symmetric twin jets).  If the jet has a luminosity of
$L\sim 10^{38}\ergps$ (comparable to the Eddington limit for $1\Msun$), the
object could appear to have a luminosity of $2\pi \gamma^2 L\sim 6\times
10^{39}\ergps$ if viewed within $20\degmark$ of its jet.  Of course the jet
could be more powerful (as indicated for GRS~1915+105; Mirabel
\& Rodriguez 1995) and the efficiency for the conversion of jet power
into X-ray flux, presumably via shocks, is unknown.  We note that this
process in blazars may be highly efficient and a hard power-law X-ray
spectrum is produced.

Theoretical models of relativistic jets have been developed in the
context of radio-loud AGN (see review by Sikora 1994 and references
within).  In the most simple models, the jet radiation is synchrotron
emission from relativistic electrons gyrating in the magnetic field of
the jet plasma.  Self-absorption and synchrotron self-Compton
processes are known to be important for understanding the observed
spectrum.  In addition, pair-production may be influential in
determining the $\gamma$-ray emission at the base of the jet.  It is
not clear how well these models for AGN jets will describe jets in
Galactic superluminal sources.  It is unlikely that a simple scaling
of the spectrum of a beamed AGN (e.g. a blasar) will successfully
describe the spectrum of such a source.  Thus, significant theoretical
work is required in order to make predictions for the full-band
spectrum of a beamed Galactic superluminal-type source.

One prediction we can make is that non-thermal continuum emission
would dominate at all wavelengths.  Thus, spectral features, such as
the fluorescent iron K$\alpha$ emission line expected from the
accretion disk, would be very weak.  This analogy with blazars might
also suggest that these objects should display extremely rapid
variability.  Future observations with the next generation of imaging
X-ray observatories (with larger collecting area) will be able to
address these issues.

The statistics for such a scheme appear to be acceptable.  The two known
Galactic superluminal objects have inclinations to our line of sight of
$\sim 70\degmark$ and $\sim 85\degmark$.  Assuming that the Galactic
superluminal sources are randomly oriented, the fact that we see a source
as close to the plane of the sky as $85\degmark$ suggests there could be at
least $\sim 10$ such sources within our Galaxy and some selection effect
could be picking out sources with high inclinations\footnote{Galactic
superluminal sources alert us to their presence by their X-ray transient
behaviour.  If the X-ray emission from such a transient event was
preferentially directed perpendicularly to the jet axis (i.e. in the plane
of the putative accretion disk), this would produce the required selection
effect.  Note that this disk-plane emission need not be directly related to
the beamed (jet) X-ray emission.}.  If there are $\sim 10$ or more such
objects per galaxy, the probability that an observer lies within the
beaming cone (taken to have a half angle of $\gamma^{-1}\sim 20\degmark$)
of one of these sources for a given galaxy can exceed 0.2.  Although the
transient nature of the Galactic superluminal sources will reduce this
probability, the existence of a steady class of jetted sources may offset
this effect.  Thus, if other galaxies harbour similar populations of beamed
sources then many of the luminous objects seen can be accounted for.

\section{Conclusions}

{\it ASCA} clearly detects X-ray emission from the nearby galaxies Dw1 and
Mf1.  The degree of absorption seen in the X-ray spectra of these two
emissions, as well as probabilistic arguments, lead us to believe that the
emission is intrinsic to the galaxy in both cases.

The Dw1 emission, Dw1-X1, is consistent with being from a point source
with a 0.5--10\,keV luminosity of $1.4\times 10^{39}\ergps$ at the
distance of Dw1.  It has a hard spectrum well described by a power-law
of photon index $\Gamma=2.0$ absorbed by a column density $N_{\rm
H}\sim 1.4\times 10^{22}\pcmsq$ of cold material.  The apparent
3\,arcmin offset (corresponding to 2.6\,kpc at the distance of Dw1) of
Dw1-X1 from the optical nucleus cannot be reasonably explained by
pointing inaccuracies.  Thus Dw1-X1 appears to be either a luminous
non-nuclear accreting X-ray source of the type found in many other
nearby galaxies, or a X-ray bright supernova remnant similar to
SN1978K in NGC~1313.  No variability is seen, although variability
would have to be dramatic in order to be detectable in such a faint
source.  No nuclear emission is seen from Dw1: the corresponding limit
on the 2--10\,keV flux of the nucleus is $F_{2-10}\approxlt 1\times
10^{-13}\ergpcmsqps$.

In contrast, the Mf1 emission is not pointlike.  Both the hard and
soft band images clearly show extension beyond the point spread
function.  The overall spectrum is hard and well described by a
power-law with $\Gamma=1.8$ modified by absorption from a cold column
density of $N_{\rm H}\sim 1\times 10^{22}\pcmsq$.  The spectral data
rule out the possibility that any significant part of the observed
emission originates from thermal ($kT\sim 1\keV$) ISM emission: an
upper limit on the 0.5--10\,keV luminosity of any such thermal
component is $1.4\times 10^{38}\ergps$ (which is 5 per cent of the
total observed X-ray emission).  We interpret this emission as being
from a number of accreting stellar systems.  We cannot distinguish
between a small number of luminous sources (each with a 0.5--10\,keV
luminosity of $\sim 10^{39}\ergps$) or a large number of ordinary
LMXB.  There might also be a LLAGN present.  Further progress requires
high spatial resolution X-ray observations.

Prompted by our results on Dw1-X1, we discuss the physical nature of
non-nuclear sources with luminosities up to $\sim 10^{40}\ergps$.
Many such sources are now known with an average of about one per
galaxy.  Given the compact and variable nature of these sources, it is
tempting to identify them as hyper-luminous XRB.  In particular, we
discuss a possible connection between such sources and Galactic
superluminal objects.  The complexities of realistic relativistic jet
models make detailed predictions difficult, but we have demonstrated
that this hypothesis is feasible on energetic grounds.

\section*{Acknowledgements}

CSR and AJL thank PPARC and ACF thanks the Royal Society for support.
WNB thanks the National Science Foundation (USA) for support.  The
POSS plate image of Mf1 was obtained via the Digitized Sky Survey
which was produced at the Space Telescope Science Institute (ST ScI)
under U.~S. Government grant NAG W-2166.

\end{document}